\documentclass[aps,amssymb,amsmath,twocolumn, pra]{revtex4}

\usepackage{amssymb}
\usepackage{amsmath}
\usepackage{draftcopy}
\usepackage{graphicx}
\usepackage{epsfig}
\usepackage{dcolumn}
\usepackage{bm}
\usepackage{color}
\usepackage{hyperref}
\usepackage[up]{subfigure}
\usepackage{natbib}

\newcommand{\be}{\begin{equation}}
\newcommand{\ee}{\end{equation}}
\newcommand{\bc}{\begin{center}}
\newcommand{\ec}{\end{center}}
\newcommand{\bea}{\begin{eqnarray}}
\newcommand{\eea}{\end{eqnarray}}
\newcommand{\ra}{\rangle}
\newcommand{\la}{\langle}

\begin{document}




\title{Quantum to classical transition and entanglement sudden
  death in Gaussian states under local heat bath dynamics}  
\author{Sandeep K \surname{Goyal}}
\email{goyal@imsc.res.in}
\affiliation{Center for Quantum Sciences, The Institute of
  Mathematical Sciences, CIT campus, Taramani, Chennai 600 113, India}

\author{Sibasish \surname{Ghosh}}
\email{sibasish@imsc.res.in}
\affiliation{Center for Quantum Sciences, The Institute of
  Mathematical Sciences, CIT campus, Taramani, Chennai 600 113, India}

\begin{abstract}
Entanglement sudden death in spatially separated two-mode Gaussian
states coupled to local thermal and squeezed thermal baths is studied
by mapping the problem to that of the quantum-to-classical
transition. Using Simon's criterion concerning the characterisation
of classicality in Gaussian states, the time to ESD is calculated by
analysing the covariance matrices of the system. The results for the
two-mode system at $T=0$ and $T>0$ for the two types of bath states
are generalised to $n$-modes, and are shown to be similar in nature to the
results for the general discrete $n$-qubit system. 
\end{abstract}
\maketitle


\section{Introduction}
Entanglement is one of the basic features that distinguish quantum
systems from their classical counterparts and is the most useful
resource in Quantum Information Theory (QIT)\cite{Nielsen}. It is
indispensible for essential quantum information tasks like quantum
computation and teleportation as well as super-dense coding and
one-way communication, to name a few. However, more often than not,
quantum systems do not operate while completely insulated from its
environment. Unfortunately, quantum systems are inherently fragile and
extremely sensitive to environmental interactions. It is a common
belief that the evolution of a  
quantum system in contact with a dissipative environment results in an
asymptotic transition to classicality \cite{Isar05, Zurek91,
  Schlosshauer} and thus an inevitable loss of 
entanglement, since classicality subsumes disentanglement. Recently 
however it has been shown \cite{Zyczkowski01, YuEbr, Roszak10} that there exists a certain
class of two-qubit states which display a finite entanglement decay
time. This phenomenon is aptly called \emph{Entanglement Sudden Death}
(ESD) and cannot be predicted from quantum decoherence which is an
asymptotic phenomenon.  

ESD has important and obvious implications for the success of quantum
information tasks. Much research has been done on ESD in discrete
quantum states. Qasimi \emph{et al.} \cite{Qasimi08} have shown that
ESD occurs for a class of two-qubit states called `$X$' states at
non-zero temperatures. The authors of the present paper have shown
that there is ESD for all $n$-qubit states at finite temperature
\cite{SKG10} and have also given a sufficiency condition for the
existence of ESD for a system consisting of $n$ number of $d$-dimensional
subsystems as well.  The study of the evolution of entanglement as
well as that of its sudden death in continuous variable systems has
also received a great attention, for example, the paper by Dodd {\em et al.}
\cite{Dodd04} deals with ESD
from the point of view of a separable representation of the joint
Wigner function of two-mode Gaussian states. Their results do not have an
explicit presence of temperature. Also, the time to ESD for the
maximally entangled EPR state is shown to be a lot smaller than that
of any other initially entangled two-mode Gaussian state. This is
counter-intuitive as one would expect the ESD time for a maximally
entangled state to be the largest. In \cite{Diosi03}, Di\'{o}si has
given a bound on the time of ESD by using a theorem on `entanglement
breaking quantum channels \cite{Shor02, Holevo99, Ruskai02}'. In
\cite{Marek08} Marek {\em et al.} have addressed a different problem. In
this paper they obtained a class of states which is tolerant against
the decoherence at zero temperature. People have studied the
decoherence in infinite dimensional systems interacting with different
kinds of bath and with different system-environment models \cite{Wilson03, Rajagopal01, 
  Rendell03, Vasile09}.


Since classicality subsumes disentanglement, the ESD problem can be
embedded into the corresponding quantum-to-classical transition problem. Thus,
the time taken for the system to attain classicality will be an upper
bound on the time at which entanglement dies. We use the criterion due to
Simon \cite{Simon00, Arvind95} for checking the classicality of Gaussian states in terms
of their covariance matrices $V$, which enables us to characterise the
quantum-to-classical transition. If $\mathbb{I}$ be the covariance
matrix corresponding to the vacuum state, then Simon's criterion
states that Gaussian states attain classicality if and only if the
difference $V-\mathbb{I}$ becomes positive semi-definite. Which is
equivalent to Sudarshan-Glauber $P$ function being positive.  We use this
condition to find out the transition time to classicality ($t_c$) and
hence the time to ESD.  Di\'{o}si {\em et al.} \cite{diosi02}
discusses about the positivity of Wigner function and $P$ function under 
Markovian evolution.  We find that the behaviour of the Gaussian
system, as regards to ESD, depends on the state and the temperature of the
bath it is coupled to. We have shown that all Gaussian states show ESD
at all temperatures $T>0$, whereas some do not at $T=0$ when coupled
to a thermal bath. However, all states again show ESD at $T=0$ when the state of
the bath is squeezed thermal. 

We begin this article by writing down the master equation for a single
harmonic oscillator (Sec.  \ref{master-equation}) in contact with a
thermal bath of infinitely many 
oscillators, taken at some finite non-zero temperature. We express the
states of this system in the Sudarshan-Glauber $P$-representation and
derive the corresponding quantum analog of the classical linear
Fokker-Planck equation (Sec. \ref{coherent-state-rep}). This
enables us to use covariance matrices to 
characterize classicality of the single-mode system in the presence of
thermal baths (Sec. \ref{covariance-matrix}). Once we have this
mechanism in place, we use it to 
determine the action of the thermal bath  (Sec.
\ref{ESD-two-mode}) and squeezed thermal bath  (Sec.
\ref{squeezed-bath}) on a two-mode Gaussian state. Using Simon's
criterion, we calculate the time to 
classicality and hence are able to determine when a system will show
ESD. Furthermore, using the two-mode analysis, we are able to predict
the possibility of ESD in any $n$-mode Gaussian state since we make the
important observation that Simon's criterion $V-\mathbb{I} \geq 0$
is independent of the number of modes of the system. Hence we are able
to generalize our earlier statements to $n$-mode systems as
well (Sec. \ref{n-mode}). Section \ref{conc} provides the conclusion of  the paper.

\section{ Master equation for single mode system}\label{master-equation} The irreversible time
evolution of the state $\rho_s(t)$ of a single harmonic oscillator
coupled to a bath is described by the Lindblad master equation (ME)
\cite{lindblad}, 

\begin{align}
\label{mastereqn}
\frac{d}{dt}\rho_s(t) &= -i\omega_0[a^{\dagger}a,\rho_s(t)]\nonumber\\
&+\gamma_0(N+1)\left\{a\rho_s(t) a^{\dagger}-\frac{1}{2}
  a^{\dagger}a\rho_s(t) -\frac{1}{2} \rho_s(t)a^{\dagger}a\right\}\nonumber\\
&+\gamma_0(N)\left\{a^{\dagger}\rho_s(t) a-\frac{1}{2}
  aa^{\dagger}\rho_s(t) -\frac{1}{2} \rho_s(t)aa^{\dagger}\right\}.
\end{align}

The initial state of the entire system is of the form
$\rho_s(0)\otimes \rho_{th}$, where $\rho_s(0)$ is the initial state
of the system and the thermal state $\rho_{th}$ is the initial state
of bath. Since the bath we are considering is very large, the state of
the bath will not change during the evolution. The first term of the ME
describes free  evolution generated by the system Hamiltonian $H =
\omega_0 a^{\dagger}a$ while the rest are interaction terms with the
bath, where the dissipative coupling is provided through the damping
rate $\gamma_0$. Here $N = (e^{\beta \omega_0}-1)^{-1}$ is the mean
number of quanta in a mode with frequency $\omega_0$. This ME can
be used to describe, for example, the damping of an electromagnetic
field inside a cavity where $a$ and $a^{\dagger}$ denote the creation
and annihilation operators of the cavity mode. The mode outside the
cavity plays the role of the environment with dissipative coupling
rate $\gamma_0$. 
 
\section{ Coherent state representation}\label{coherent-state-rep} We can transform the master
equation described by (\ref{mastereqn}) into a continuous diffusion
process by using the coherent state representation, 
\begin{align}\label{P-rep}
\rho_s(t) &= \int d^2\alpha P(\alpha, \alpha^*,t)|\alpha\ra\la\alpha|,
\end{align}
where $P(\alpha, \alpha^*,t)$ is the Sudarshan-Glauber $P$-function
\cite{Glauber63, Sudarshan63} and the integration in (\ref{P-rep}) is
over the entire complex plain. The quasiprobability distribution,
called thus because $P(\alpha, \alpha^*,t)$ can take negative values
for some $\alpha$, satisfies the normalization 
\begin{align}
{\rm tr}_s\rho_s(t) &= \int d^2\alpha  P(\alpha, \alpha^*,t) =1.
\end{align}
Substituting (\ref{P-rep}) into (\ref{mastereqn}) and using the properties
\begin{align}\label{eqn4}
a|\alpha\ra\la \alpha | &= \alpha|\alpha\ra\la \alpha |,\nonumber\\
a^{\dagger}|\alpha\ra\la \alpha | &=\left( \frac{\partial}{\partial \alpha}+
\alpha^*\right)|\alpha\ra\la \alpha |,
\end{align}
we get the following equation for the evolution of $P(\alpha, \alpha^*,t)$:
\begin{widetext}
\begin{align}\label{P-master}
\frac{\partial}{\partial t}P(\alpha, \alpha^*,t) &= -\left[
  \left(-i\omega_0-\frac{\gamma_0}{2}\right)\frac{\partial}{\partial
    \alpha}\alpha+
  \left(i\omega_0-\frac{\gamma_0}{2}\right)\frac{\partial}{\partial
    \alpha^*}\alpha^*\right] P(\alpha, \alpha^*,t) +
\gamma_0N\frac{\partial^2}{\partial\alpha\partial\alpha^*}P(\alpha,
\alpha^*,t).
\end{align}
\end{widetext}
This is structurally similar to the classical linear Fokker-Planck equation \cite{Risken} and can be solved using the Gaussian ansatz
\begin{align}
P(\alpha, \alpha^*,t) &= \frac{1}{\pi
  \sigma^2(t)}\exp\left[-\frac{|\alpha-\beta(t)|^2}{\sigma^2(t)}\right],
\end{align}
given the initial condition $P(\alpha, \alpha^*,0) = \delta^2(\alpha-\alpha_0)$, where $|\alpha_0\ra$ is the initial coherent state. Here the mean amplitude is given by $\beta(t) = \int d^2\alpha\alpha P(\alpha, \alpha^*,t) = \alpha_0e^{(-i\omega_0-\gamma_0/2)t}$ and the variance is $\sigma^2(t) =  N(1-e^{-\gamma_0t})$. Since we are interested in dissipation, we choose to ignore the $\omega_0$ term which only contributes to free evolution. The linearity of the evolution map ensures that any density matrix $\rho(0) = \int P(\lambda, \lambda^*,0)|\lambda\ra \la \lambda|d^2\lambda$ will evolve to $\rho(t) = \int P(\alpha, \alpha^*, t)|\alpha\ra\la\alpha|d^2\alpha$, where $P(\alpha,\alpha^*, t) = \int
P(\lambda, \lambda^*,0)\exp\left(-\frac{|\alpha-\beta(t)|^2}{\sigma^2(t)}\right)d^2\beta$ with $\beta = \lambda e^{-\gamma_0t/2}$. Therefore, thermal evolution of the $P$-distribution manifests as a convolution of the $P$ and the thermal distribution $P_{th} = \exp(-|\alpha|^2/\sigma^2(t))$. 

\section{Covariance matrix}\label{covariance-matrix}
Our aim is to express the evolution of the state in terms of the covariance matrix, and to that end we write down the symmetric characteristic function $\chi(\alpha,t)$ \cite{Cahill69A, Cahill69B}
\begin{align}\label{characteristic}
\chi(\alpha, t) &= e^{-\frac{|\alpha|^2}{2}}\mathcal{F}[P(\alpha, \alpha^*,
  t)] = \exp\left(-\frac{X^TVX}{4}\right),
\end{align}
where $\mathcal{F}$ denotes the Fourier transform and $V$ is the covariance matrix. The vector $X^T = (q,~p)$ where $q = \frac{1}{\sqrt{2}} (\alpha+\alpha^{\ast})$ and $p = \frac{i}{\sqrt{2}} (\alpha^{\ast}-\alpha)$ are the position and momentum variables. The evolution of $\chi(\alpha,t)$ follows from that of $P(\alpha, \alpha^*,t)$ (equation \ref{P-master}). We have, from (\ref{characteristic})
\begin{align}
\chi(\alpha, t)e^{\frac{|\alpha|^2}{2}} &= \chi(\beta e^{-\frac{\gamma_0}{2}t},0)e^{-\frac{|\alpha|^2 e^{-\gamma_0t}}{2}}  e^{-\frac{|\alpha|^2\sigma^2(t)}{4}}.
\end{align}
The covariance matrix $V$ can be written as:
\begin{align}
V_{ij} &= \int (XX^T)_{ij} \mathcal{F}^{-1}\chi(\alpha,t)d^2qd^2p,
\end{align}
and hence
\begin{align}\label{covariance-evolution}
V(t) & = e^{-\gamma_0t}V(0) + \left(\frac{N}{2}+1\right)(1-e^{-\gamma_0t})\mathbb{I}.
\end{align}

A \emph{Gaussian channel} is a map that takes Gaussian states to
Gaussian states, an example of which is the evolution, given by
Eq. (\ref{mastereqn}) for a harmonic
oscillator  coupled with a thermal bath. The
evolution of the covariance matrix $V$ of the system, under the action
of a general Gaussian channel, can be characterized by two matrices
$A$ and $B$: 
\begin{align}\label{eqn10}
V_f &= AVA^T + B,
\end{align}
where $B$ is a positive operator \cite{Heinosaari09}. For a thermal
bath, we have from (\ref{covariance-evolution}), $A =
e^{-\gamma_0t/2}\mathbb{I}$ and $B =
\left(\frac{N}{2}+1\right)(1-e^{-\gamma_0t})\mathbb{I}$. We can hence
characterize the action of a  thermal bath completely using these two
matrices. The characterization, given by Eq. (\ref{eqn10}),
guarantees that $V(t)$ of Eq. (\ref{covariance-evolution}) is a
bona fide covariance matrix for all finite time $t$.

\section{ESD of two-mode Gaussian state}\label{ESD-two-mode}
Consider a two-mode system coupled to two identical local thermal
baths. Let us assume that the initial two-mode ($4 \times 4$)
covariance matrix $V_0$ represents an entangled state. Its subsequent
evolution is given by 
\begin{align}
\label{evolved_covariance}
V(t) &= (A\oplus A) V(0)(A\oplus A)^T + (B \oplus B)\nonumber\\
&=e^{-\gamma_0t}V(0) + \left(\frac{N}{2}+1\right)(1-e^{-\gamma_0t})\mathbb{I}.
\end{align}
where $A = e^{-\gamma_0t/2}\mathbb{I}_4$ and $B =
\left(\frac{N}{2}+1\right)(1-e^{-\gamma_0t})\mathbb{I}_4$. From a
quantum optics point of view, we know \cite{Simon00}, in view of
the $P$-representation, that the state $\rho = \int d^2\alpha d^2\beta
P(\alpha, \beta) |\alpha\ra\la\alpha|\otimes|\beta\ra\la\beta|$ is
classical only when $P(\alpha,\beta)$ is non-negative for all $\alpha$
and $\beta$. This interpretaion of classicality can be translated into
the language of the covariance matrix $V(t)$. Thus, a two-mode Gaussian
state will be classical at some time $t$ if and only if  $V(t) \geq \mathbb{I}$
\cite{Simon00}. Clearly, being entangled, the initial covariance
matrix satisfies: $V(0) < \mathbb{I}$. However, since the evolution
(\ref{evolved_covariance}) is dissipative, the system will attain
classicality after a time $t_c$ so that $V(t_c) \geq \mathbb{I}$. This
condition is equivalent to $n_{min}(t_c) \geq 1$, where $n_{min}(t_c)$
is the smallest eigenvalue of $V(t_c)$. Using
(\ref{evolved_covariance}), it can be written as the evolution of
$n_{min}(0)$ which is the smallest eigenvalue of $V_0$. Since $V(0)$
is not classical, $n_{min}(0)<1$. 
\begin{align}
\label{n_min}
n_{min}(t_c) &= e^{-\gamma_0t_c}n_{min}(0)+ \left(\frac{N}{2}+1\right)(1-e^{-\gamma_0t_c}).
\end{align}
Here we find that $n_{min}(t_c) \geq 1$ always for $t_c \ge
-\frac{1}{\gamma_0}\ln\left(\frac{N}{N+2-2n_{min}(0)}\right)$. An
appropriate choice of $V_0$ allows us to make $n_{min}(0)$ arbitrarily
small (which is the case in EPR states, which are maximally entangled initially) and thus get an upper bound on
the transition time $t_c$, given 
by $t_{max} =-\frac{1}{\gamma_0}\ln\left(\frac{N}{N+2}\right)$. For
$T>0$, we have $N>0$ and hence $t_{max}$ is finite and non-zero. This
proves that there is always ESD at non-zero temperatures. However,
when $T=0$, \emph{i.e.} $N=0$, we have $t_{max}\to \infty$ and hence
no quantum-to-classical transition is seen at finite times. This does not,
however, rule out ESD since non-classicality does not necessarily
imply that there is entanglement. 

For $T=0$, let us consider a particular form of the initial covariance matrix representing a symmetric two-mode Gaussian state.
\begin{align}
V(0) &= \left(\begin{array}{cccc}
n&0&k_x&0\\
0&n&0&-k_y\\
k_x&0&n&0\\
0&-k_y&0&n
\end{array}\right).
\label{particular_V0}
\end{align}
Here $k_x$ and $k_y$ are positive and satisfy $n^2 -
(max\{k_x,k_y\})^2\ge 1$ (in order that $V(0)$ is a bona fide
covariance matrix). This covariance matrix represents an
entangled state when $(n-k_x)(n-k_y)<1$ \cite{EOF03}. Under evolution
(\ref{evolved_covariance}) and taking $k_x=k_y$, we have $n(t) =
ne^{-\gamma_0t}+(1-e^{-\gamma_0t})$ and $k_x(t) =
k_xe^{-\gamma_0t}$. At the quantum-to-classical transition time $t_c$,
the states become separable: $(n(t_c)-k_x(t_c))^2 \ge 1$. This
condition is, in this case, equivalent to $n(t_c)-k_x(t_c) \ge
1$. However, we have verified that there is no positive finite value
of $t_c$ that satisfies this condition. Hence, we see that for the
zero-temperature case, there are states given by
(\ref{particular_V0}) that do not show ESD. This can be easily
generalized to the case when $k_x \ne k_y$. This result is similar to
the two-qubit case where we see that ESD occurs for all states at
non-zero temperature. For $T = 0$, there exist states which do not show
ESD \cite{SKG10}.  

\section{ Squeezed thermal bath}\label{squeezed-bath} If the initial
state of the bath is a squeezed 
thermal state, then the master equation  (\ref{mastereqn})  is replaced by \cite{Ferraro05}:  
\begin{align}
\label{mastereqn2}
\frac{d}{dt}\rho_s(t) &= -i\omega_0[a^{\dagger}a,\rho_s(t)]\nonumber\\
&+\gamma_0(N+1)\left\{a\rho_s(t) a^{\dagger}-\frac{1}{2}
  a^{\dagger}a\rho_s(t) -\frac{1}{2} \rho_s(t)a^{\dagger}a\right\}\nonumber\\
&+\gamma_0N\left\{a^{\dagger}\rho_s(t) a-\frac{1}{2}
  aa^{\dagger}\rho_s(t) -\frac{1}{2}
  \rho_s(t)aa^{\dagger}\right\}\nonumber\\
&-\frac{\gamma_0}{2}M^* \left\{2a\rho a -a^2\rho - \rho a^2\right\}\nonumber\\
&-\frac{\gamma_0}{2}M \left\{2a^{\dagger}\rho a^{\dagger}
  -\left(a^{\dagger}\right)^2\rho - \rho
  \left(a^{\dagger}\right)^2\right\},
\end{align}
where $M = - \frac{1}{2}\sinh(2r)e^{i\phi}(2N_{th} +1)$ and $2N+1 = \cosh(2r)(2N_{th}+1)$. The quantity $N_{th}$ is the average number of photon in the thermal state and $r$ and $\phi$ are the squeezing parameters. Repeating the same procedure as earlier, we finally write down the evolution in terms of covariance matrices:
\begin{align}
V(t) &= e^{-\gamma_0t}V(0) + (1-e^{-\gamma_0t})V_{\infty},
\end{align}
where $V_{\infty}$ is given by
\begin{align}
V_{\infty} &= \left(\begin{array}{cc}
\frac{N}{2}+1 + Re\{M\}& Im\{M\}\\
Im\{M\} & \frac{N}{2}+1 - Re\{M\}
\end{array}\right).
\end{align}
In this case, for a squeezed thermal bath, $A =
e^{-\gamma_0t}\mathbb{I}$ and $B =(1-e^{-\gamma_0t})V_{\infty}$. If
there is no squeezing (\emph{i.e.} if we set $M=0$ and $N=N_{th}$), we
get $V_{\infty}=(N/2+1)\mathbb{I}$ and thus recover the unsqueezed
result namely, Eq.  (\ref{evolved_covariance}). 

Thus the evolution of the two-mode Gaussian state can be written as
\begin{align}
V(t)& = e^{-\gamma_0t}V(0) + (1-e^{-\gamma_0t})(V_{\infty}\oplus V_{\infty}).
\end{align}
If the smallest eigenvalue of $V(0)$ is $n(0)$ then the classicality condition $V(t) \ge \mathbb{I}$ can be written as \cite{Horn-Johnson}:
\begin{align}
n(t) \ge e^{-\gamma_0t}n(0) + (1-e^{-\gamma_0t})\left(\frac{N}{2}+ 1 - |M|\right) \ge 1,\label{class-sq}\\
{\rm i.e,}t \ge -\frac{1}{\gamma_0}\ln\left(\frac{N - 2|M|}{N + 2 -2|M| - 2n(0)}\right),
\end{align}
where in Eq.  (\ref{class-sq}), $n(t)$ is the smallest eigenvalue of
$V(t)$ and $\left(\frac{N}{2} +1 -|M|\right)$ is the smallest
eigenvalue of $V_{\infty}$.
Since $n(0)<1$, $t_c$ is finite and positive. Therefore, the
transition to classicality and thus ESD is ensured for the squeezed
thermal bath for non-zero temperatures. However, unlike the thermal
bath case, we see here that $N$ does not become zero with temperature
$T$ and hence a quantum-to-classical transition always happens at zero
temperature for the squeezed thermal bath which ensures ESD.

\section{ESD for $n$-mode Gaussian states}\label{n-mode}
In this section we generalize the result mentioned previously for $n$-mode Gaussian
states in the presence of local thermal and squeezed thermal
baths. Consider a covariance matrix $V_n(0)$ corresponding to an
$n$-mode Gaussian state. Let the matrices $A$ and $B$ characterise
the single mode Gaussian channel for a  given bath. The
evolution of the covariance matrix can be written as [as a
generalization of Eq.  (\ref{evolved_covariance} )],
\begin{align}
V_n(t) =& (A\oplus A\oplus \cdots \oplus A)V_n(0)(A\oplus A\oplus
\cdots \oplus A)^T \nonumber\\
&+ (B\oplus B \oplus \cdots \oplus B),
\end{align}
If $n(0)$ is the smallest eigenvalue of $V_n(0)$, the classicality
condition $V_n(t) \ge \mathbb{I}$ gives rise to 
an equation   which is same as Eq.  (\ref{n_min}) (
Eq.  (\ref{class-sq})) representing $n_{min}(t)$($n(t)$) as the smallest
eigenvalue 
of $V_n(t)$  in the case of the thermal (squeezed thermal)
bath. Therefore, all $n$-mode Gaussian states show
quantum-to-classical transition at finite temperature.


\section{ Conclusion}\label{conc} In this article we have analysed the 
phenomenon of entanglement sudden death for the case of Gaussian
states coupled to local identical thermal and squeezed thermal
baths. We have mapped this problem of non-asymptotic decay of
entanglement to the quantum-to-classical transition phenomenon by
noting that classicality subsumes separability. Using the powerful
criterion given by Simon, we have characterized the classicality of
the Gaussian states and calculated the time taken to attain
classicality. We have shown whether ESD is possible at zero
temperature as well as at non-zero temperatures for the two different
types of baths considered. We have been able to generalize our two-mode
result to a general $n$-mode system by exploiting the
mode-independence of Simon's criterion. Finally, we have noted the
similarity that the ESD results for Gaussian systems have
with discrete system results \cite{SKG10}. We believe that the same might hold for
non-Gaussian states as well, work on which is currently in progress. 

\section*{Acknowledgement}
We would like to thank L. Di\'{o}si, A. Rajagopal, R. Horodecki,
M. S. Kim and J.-H. An for their helpful suggestions and comments.


\begin{thebibliography}{10}

\bibitem{Nielsen}
M.~A. Nielsen, I.~L. Chuang, {\it Quantum Computation and Quantum
  Information\/} (Cambridge University Press, Cambridge, 2000).

\bibitem{Isar05}
A.~Isar, {\it Rom. Jour. Phys.\/} {\bf 50}, 147 (2005).

\bibitem{Zurek91}
W.~H. Zurek, {\it Physics today\/} {\bf 44(10)}, 36 (1991).

\bibitem{Schlosshauer}
M.~A. Schlosshauer, {\it Decoherence and the quantum-to-classical transition\/}
  (Springer, 2007).

\bibitem{Zyczkowski01}
K.~Zyczkowski, P.~Horodecki, M.~Horodecki, R.~Horodecki, {\it Phys. Rev. A\/}
  {\bf 65}, 012101 (2001).

\bibitem{YuEbr}
T.Yu, J.~Eberly, {\it Phys. Rev. Lett.\/} {\bf 93}, 140404 (2004).

\bibitem{Roszak10}
K.~Roszak, P.~Horodecki, R.~Horodecki, {\it Phys. Rev. A\/} {\bf 81}, 042308
  (2010).

\bibitem{Qasimi08}
A.~Al-Qasimi, D.~F.~V. James, {\it Phys. Rev. A\/} {\bf 77}, 012117 (2008).

\bibitem{SKG10}
S.~K. Goyal, S.~Ghosh, {\it arxiv:1003.1248\/}  (2010).

\bibitem{Dodd04}
P.~J. Dodd, J.~J. Halliwell, {\it Phys. Rev. A\/} {\bf 69}, 052105 (2004).

\bibitem{Diosi03}
L.~Di\'{o}si, {\it Lect. Notes Phys.\/} {\bf 622}, 157 (2003).

\bibitem{Shor02}
P.~W. Shor, {\it J. Math. Phys.\/} {\bf 43}, 4334 (2002).

\bibitem{Holevo99}
A.~S. Holevo, {\it Russian Math. Surveys\/} {\bf 53}, 1295 (1999).

\bibitem{Ruskai02}
M. Horodecki, P. W. Shor, and M. B. Ruskai, {\it Rev. Math. Phys.} {\bf 15}, 629 (2003).

\bibitem{Marek08}
P.~Marek, J.~Lee, M.~S. Kim, {\it Phys Rev A\/} {\bf 77}, 032302 (2008).

\bibitem{Wilson03}
D.~Wilson, J.~Lee, M.~S. Kim, {\it J. Mod. Opt.\/} {\bf 50}, 1809 (2003).

\bibitem{Rajagopal01}
A.~K. Rajagopal, R.~W. Rendell, {\it Phys. Rev. A\/} {\bf 63}, 022116 (2001).

\bibitem{Rendell03}
R.~W. Rendell, A.~K. Rajagopal, {\it Phys. Rev. A\/} {\bf 67}, 062110 (2003).

\bibitem{Vasile09}
R.~Vasile, S.~Olivares, M.~G.~A. Paris, S.~Maniscalco, {\it Phys. Rev. A\/}
  {\bf 80}, 062324 (2009).

\bibitem{Simon00}
R.~Simon, {\it Phys. Rev. Lett.\/} {\bf 84}, 2726 (2000).

\bibitem{Arvind95}
Arvind, B. Dutta, N. Mukunda, and R. Simon, {\it Pramana -- journal of physics} {\bf 45}, 471 (1995).

\bibitem{diosi02}
L.~Di\'{o}si, C.~Kiefer, {\it J. Phys. A: Math. Gen.\/} {\bf 35}, 2675 (2002).

\bibitem{lindblad}
G.~Lindblad, {\it Commun. Math. Phys.\/} {\bf 48}, 119 (1976).

\bibitem{Glauber63}
R.~J. Glauber, {\it Phys. Rev.\/} {\bf 131}, 2766 (1963).

\bibitem{Sudarshan63}
E.~C.~G. Sudarshan, {\it Phys. Rev. Lett.\/} {\bf 10}, 277 (1963).

\bibitem{Risken}
H.~Risken, {\it The Fokker-Planck equation -- Methods of solution and
  applications\/} (Springer, 1989).

\bibitem{Cahill69A}
K.~E. Cahill, R.~J. Glauber, {\it Phys. Rev.\/} {\bf 177}, 1857 (1969).

\bibitem{Cahill69B}
K.~E. Cahill, R.~J. Glauber, {\it Phys. Rev.\/} {\bf 177}, 1882 (1969).

\bibitem{Heinosaari09}
T. Heinosaari, A. S. Holevo, and M. M. Wolf, {\it Quantum Inf. Comp.} {\bf 10}, 0619 (2010).

\bibitem{EOF03}
G.~Giedke, M.~M. Wolf, O.~Kruger, R.~F. Werner, J.~I. Cirac, {\it Phys. Rev.
  Lett.\/} {\bf 91}, 107901 (2003).

\bibitem{Ferraro05}
A. Ferraro, S. Olivares, and M. G. A. Paris, {\it Gaussian states in quantum information}, Lecture notes, ISBN 88-7088-483-X (Biliopolis, Napoli, 2005).
  (2005).

\bibitem{Horn-Johnson}
R.~A. Horn, C.~R. Johnson, {\it Matrix Analysis\/} (Cambridge University Press,
  1985) {\em page 194} (Theorem--4.3.27).

\end{thebibliography}

\end{document}